# Subdivisions and Crossroads: Identifying Hidden Community Structures in a Data Archive's Citation Network

Sara Lafia (0000-0002-5896-7295)*, Lizhou Fan (0000-0002-7962-9113)†, Andrea Thomer (0000-0001-6238-3498)†, Libby Hemphill (0000-0002-3793-7281)*†

* ICPSR, University of Michigan, Ann Arbor, MI, USA
† School of Information, University of Michigan, Ann Arbor, MI, USA

Email: slafia@umich.edu, lizhouf@umich.edu, athomer@umich.edu, libbyh@umich.edu

**Abstract**
Data archives are an important source of high-quality data in many fields, making them ideal sites to study data reuse. By studying data reuse through citation networks, we are able to learn how hidden research communities – those that use the same scientific datasets – are organized. This paper analyzes the community structure of an authoritative network of datasets cited in academic publications, which have been collected by a large, social science data archive: the Interuniversity Consortium for Political and Social Research (ICPSR). Through network analysis, we identified communities of social science datasets and fields of research connected through shared data use. We argue that communities of exclusive data reuse form "subdivisions" that contain valuable disciplinary resources, while datasets at a "crossroads" broadly connect research communities. Our research reveals the hidden structure of data reuse and demonstrates how interdisciplinary research communities organize around datasets as shared scientific inputs. These findings contribute new ways of describing scientific communities in order to understand the impacts of research data reuse.

**Keywords**
archival science, community detection, data citation, data reuse, network analysis

# Introduction

Data are essential resources for social science research, and data creators' contributions should be rewarded (Alter & Gonzalez, 2018). In addition to ensuring credit, measures of data reuse such as downloads and citations can reveal a dataset's role in a research community and provide insights into how researchers engage with data (Cousijn et al., 2019). Analyzing data citations reveals data citation practices and provides a way to quantify the analytical utility and disciplinary reach of data collections (Buneman et al., 2021). However, it has typically been challenging to find these measures because download data is not widely available, and researchers inconsistently cite data (Buneman et al., 2020; Lowenberg et al., 2019). Incomplete or opaque research data citations fail to include persistent identifiers, which create obstacles to tracking data use and fail to give appropriate credit to data creators (Moss & Lyle, 2018).

Data archives – particularly domain-specific archives with robust curation services – are ideal sites to study data reuse. They provide data services that make reuse easier, making them sites of research convergence. Archives anticipate datasets that have high analytical potential for long-term preservation and community impact as “topical collections" (Fenlon, 2017; Palmer et al., 2011). Additionally, some maintain bibliographies of papers that reuse data from the archive, therefore tracking “citations” even when they are not formally included in a paper (e.g., NASA's Data Archive Centers: DAACs[1]; biodiversity data aggregators like Global Biodiversity Information Facility: GBIF[2]; and Data Observation Network for Earth: DataONE[3]). There has been relatively little scientometric research on the nature of this reuse, however.

Citations of data in these archives create networks of datasets withattributes that help us understand data reuse and its implications. For instance, understanding the context of data discovery and reuse may help us understand the distribution of ideas or topics within and between research domains, and identify datasets that exhibit exceptional long-term analytical potential (Palmer et al., 2011). Like “hibernators” among research papers (Hu & Rousseau, 2019), valuable datasets may lay dormant for years until they are discovered and “awakened” through reuse. Identifying different functions that data serve within knowledge communities can help us ensure data creators receive appropriate credit for their contributions.

Additionally, looking for new patterns of data co-use or reuse would help identify hidden communities that use archived data in novel ways. Data reuse can be viewed as an indirect form of cooperation and collaboration between researchers (Sands et al., 2012; Thomer et al., 2018; Zimmerman, 2008). Data archives promote research by providing access to datasets, and some of these datasets function as “boundary objects" (Star & Griesemer, 1989) or parts of shared information spaces (Bannon & Schmidt, 1989). Revealing hidden reuse communities and their structures helps us understand what roles data play in knowledge production and how they function as boundary objects between fields of research.

Despite recent data-sharing mandates, securing data deposits is still a challenge for data archives. Researchers are often wary of sharing data because they fear being “scooped” or are unsure how other researchers might use their data (Borgman et al., 2018; Cragin et al., 2010). Mapping the network of data citations provides evidence of data reuse that will help data producers and archives better assess the collaborative utility of data and demonstrate different types of secondary use to researchers and potential depositors.

[1] https://lpdaac.usgs.gov/resources/publications/
[2] https://www.gbif.org/resource/search?contentType=literature&relevance=GBIF_USED
[3] https://search.dataone.org/profile

In this paper, we inspect an authoritative bibliography of social science datasets cited in academic publications from the ICPSR Bibliography of Data-related Literature.[4] Specifically, we analyzed its citation graph to uncover hidden community structures and identified different roles datasets play in networked communities. By linking citations to metadata from a scholarly database, Dimensions, we were able to include attributes such as "fields of research"[5] in our analysis (Hook et al., 2018). We then used community detection algorithms to identify hidden communities within the network of data citations and identified two types of datasets that unite scientists involved in social science knowledge production: subdivision datasets and crossroads datasets. Subdivisions exclusively function as disciplinary resources used by a narrow set of fields. Crossroads, by contrast, integrate interdisciplinary research. The network structures we identify and name acknowledge the variation in reuse and help us recognize the myriad functions that datasets serve in scientific communities.

# Background and related work

## Data archives as a site for understanding scholarly communication practices

Data archives support data-intensive research by providing long-term data stewardship, access, and high-quality data curation. Notable examples of data archives with high levels of curation include GenBank, a rich repository of genetic sequence data; SESAR, a repository of metadata describing physical samples in the earth sciences, as well as links to derived datasets; and PANGAEA, a publisher for georeferenced datasets linked to earth system studies. Data sharing through archives enables researchers to find and reuse data that they did not collect. In other words, data created for one purpose can be used by new audiences to answer new questions (Brown, 2003; Wilkinson et al., 2016). Researchers can use existing data to validate previous findings, extend their data collections, or form the basis for new studies via integration or independent reuse (Gregory et al., 2020; King, 1995; Pasquetto et al., 2017; Thomer, 2022). Additionally, as more funders and journals mandate that data from grants and papers be shared openly, data archives are only growing in importance as sites of scholarly communication.

The data held in these repositories often have untapped reuse potential across disciplinary boundaries (Hey et al., 2009; Palmer et al., 2011). Such interdisciplinary research using archived data can lead to breakthrough discoveries (National Academy of Sciences, 2005; Tenopir et al., 2011). Fields of research may share an interest in explaining different aspects of the same phenomenon, giving rise to interfield theories that bridge fields of science (Darden & Maull, 1977). "Borderland disciplines" sometimes form where fields of research collide over shared resources, such as instruments or data, leading to the evolution of new techniques

[4] https://www.icpsr.umich.edu/web/pages/ICPSR/citations/
[5] According to Dimensions, the Fields of Research (FOR) is a hierarchical classification applicable for categorizing all research and development activity.

(Gökalp, 1987). Datasets that facilitate interactions between research areas therefore function as “boundary objects,” carrying multivalent analytical potential across research communities (Star & Griesemer, 1989) and facilitating knowledge exchange across boundaries. However, there has been little research on the prevalence of such datasets-as-boundary-objects. We know little about which features of datasets promote boundary crossing, or how to measure their collaborative potential.

## Data citation standards and emerging data citation networks

One way of exploring interdisciplinary data reuse – and therefore, the extent to which datasts function as boundary objects between communities – is by studying data citation networks. Efforts to promote data citation over the last 20 years have led to the adoption of new data citation practices in many communities. Milestones formalizing data citation include the Joint Declaration of Data Citation Principles (Data Citation Synthesis Group, 2014), Data Citation Roadmap for Scholarly Data Repositories (Fenner et al., 2016), and Data Citation Roadmap for Scientific Publishers (Cousijn et al., 2017). Data citation counts provide a foundation for studying the scholarly impact of scientific data and the value of data curation efforts.

The adoption of data citation principles makes it possible to analyze emerging data reuse behavior and structures of hidden research communities in data citation networks. Citation networks generally represent documents as vertices and citations of one document by another as edges (Leicht et al., 2007). Citation networks can highlight central nodes like influential institutions; heavy edges between nodes indicate important connections and processes, like the diffusion of ideas (Chen, 2017). Prior studies of citation networks have provided insights into ties between individual researchers and collaborations between research disciplines (Tomasello et al., 2017). Studies of publication citation networks (e.g., papers or journals) have also identified novel papers, measured the impact of papers and their authors, and attributed discoveries to authors (Newman, 2004).

Whereas publication citations broadly enable lineage retrieval for ideas, data citations indicate the origins and processing history of the datasets that have been used in an analysis (Bose & Frew, 2005). Data citation networks reflect connections between disciplinary literature and the research data that they draw from. They reveal the reach of research data and support the computation of bibliometrics that show the relationships and impacts of scientific products (Buneman et al., 2021). The interactional context of data production and citation also reflects relationships between data producers and consumers in a broader data economy (Vertesi & Dourish, 2011). To tap the potential of shared datasets, we examine the role that data citations play in the production and dissemination of knowledge in the social sciences.

## Exclusive and inclusive communities in knowledge organizations

The analysis of citation networks can reveal hidden organizational structures. Co-citation analysis studies the structure of science and the emergence of specialities in bibliometric networks by examining how frequently pairs of documents are invoked (Small, 1973). Author co-citation analysis reveals individual contributions to speciality areas and paradigm shifts in the research landscape (White & McCain, 1998). Citation analysis can be used to identify exclusionary community structures, such as “invisible colleges” (Price & Beaver, 1966) – in-groups that control scientific discourse, which are defined by strong ties and informal communication (Crane, 1977). Similar analyses can also detect “citation cartels" of authors who cite each other exclusively, and effectively shut out other authors who work on the same subject (Franck, 1999). In addition to exclusionary practices, citation analysis can also identify convergence in research communities. Studies of cross-field citation networks have found that fields of science tend to become more integrated, rather than exclusive, over long periods of time (Varga, 2019), albeit incrementally across neighboring disciplines (Porter & Rafols, 2009).

While the notion of “community” is central to these analytical methods, it is a difficult concept to operationalize (Orthia et al., 2021); communities may take many forms, and may play many roles. Identifying communities via data citation is further complicated by the interdisciplinary nature of data analysis and citation (Heidorn, 2008). However, we take inspiration from prior work showing that data reuse can be viewed as an indirect form of cooperation and collaboration between researchers – and groups that commonly reuse the same data might be considered communities-at-a-distance (Sands et al., 2012; Thomer et al., 2018; Zimmerman, 2008). Research data is a primary input for scientific knowledge production, making data archives important sites for identifying nascent research communities. We use community detection to reveal patterns of data reuse and examine the structure of research communities that use data as shared scientific inputs.

# Data and methods

We analyzed the ICPSR Bibliography, an authoritative source of high quality, manually-curated links between 8,071 social science studies and 101,674 publications that have cited them. At ICPSR, each study consists of one or more data files and metadata. **Table 1** provides an example of available metadata for a highly cited ICPSR study.

**Table 1**. Example of available metadata for an ICPSR Study

| Study name | Series title | Release | Citations | Subject Terms |
|---|---|---|---|---|
| Monitoring the Future: A Continuing Study of American Youth (12th-Grade Survey), 1996 | Monitoring the Future (MTF) Public-Use Cross-Sectional Datasets | 1998-10-05 | 251 | attitudes, demographic characteristics, drug use, family life, high school students, life plans, lifestyles, social behavior, social change, values, youths |

Curation of the ICPSR Bibliography is labor intensive and so the current coverage of the ICPSR Bibliography is uneven[6]. Bibliography staff search broadly for academic literature that references ICPSR studies and add literature to the Bibliography only if it analyzes ICPSR data or includes an extensive discussion of data-related methodology. Publications in the Bibliography are a mixture of materials published by the original data creator and publications that analyze existing data. The majority of materials are journal articles, reports, conference proceedings, theses, books, and book chapters. We restricted our analysis to materials published since the inception of ICPSR as an archive in 1962.

We analyzed citations for all of ICPSR's currently available studies. Many ICPSR studies have institutional principal investigators (PIs) including U.S. government agencies (e.g., U.S. Census Bureau, Department of Justice, Department of Education, Department of Health and Human Services), news outlets (CBS News, the New York Times), and university research centers (e.g., University of Michigan's Survey Research Center). Teams of individual researchers also deposit data with ICPSR. Studies in our analysis included both restricted and public data files. The terms of use for restricted data prohibit linking it to other data, so studies that include restricted data may be undercounted in terms of their potential use.

The majority of ICPSR's studies (62%) are also part of a series, meaning that they are part of a recurring collection with new data archived over time (e.g., repeated cross-sectional studies or longitudinal studies). ICPSR provides access to 278 series. We used a natural breaks classification (Jenks, 1963) to find highly cited series, which are reported in **Table 2**.

**Table 2**. Features of highly cited ICPSR series data

| Series title | Lead investigators | Studies in series | Combined citations |
|---|---|---|---|
| American National Election Study (ANES) Series | Warren E. Miller et al. and the National Election Studies | 92 | 16,771 |
| Uniform Crime Reporting Program Data Series | Federal Bureau of Investigation | 263 | 13,041 |
| Monitoring the Future (MTF) Public-Use Cross-Sectional Datasets | Lloyd D. Johnston et al. | 76 | 11,808 |
| Current Population Survey Series | U.S. Bureau of the Census | 296 | 11,012 |
| National Health and Nutrition Examination Survey (NHANES) | Kathleen Mullan Harris et al. | 3 | 6,951 |

[6] The process of retrieving citations for all studies is ongoing. Because staff are actively searching for publications that reference ICPSR datasets, these measures are minimum counts, which likely underestimate the number of papers and their relationships.

| | | | |
|---|---|---|---|
| and Followup Series | | | |
| National Survey on Drug Use and Health (NSDUH) Series | United States Department of Health and Human Services; National Institutes of Health; National Institute on Drug Abuse | 29 | 5,893 |
| National Electronic Injury Surveillance System (NEISS) Series | United States Department of Health and Human Services; Centers for Disease Control and Prevention; National Center for Injury Prevention and Control | 38 | 5,255 |
| National Crime Victimization Survey (NCVS) Series | Bureau of Justice Statistics | 85 | 4,472 |

## Network definitions

We constructed citation networks from the ICPSR Bibliography, which are summarized in **Table 3**. Given that studies from the same series have been created intentionally to be analyzed together (e.g., across years), we grouped studies by their series and referred to the resulting unit as a "dataset" – either one series with multiple studies or one study that is not part of a series. Grouping studies into ICPSR-defined series allowed us to distinguish data that were *designed* to be used together (e.g., by their project sponsor, funder, archive) from data that have been *discovered* to be useful together (e.g., by researchers who co-cite them in literature).

**Table 3**. Summary of network definitions and metrics

| ***Network*** | ***B*** | ***S*** | ***F*** |
|---|---|---|---|
| **Nodes** | publications, datasets | datasets | fields of research |
| **Edges** | Publication cited ICPSR dataset | ICPSR datasets cited in the same publication | Publication tagged with both fields cited one ICPSR study |
| **N (nodes)** | 90,922 publications; 3,363 datasets | 998 datasets | 129 research fields |
| **N (edges)** | 102,580 | 3,208 | 4,238 |
| **Node size** | constant | constant | $\log(N_{papers})$ |
| **Edge weight** | n/a | 1 for each publication in which the pair of | 1 for each ICPSR study a publication |

| | | ICPSR datasets is cited | cites |
|---|---|---|---|
| **Components** | 1,687 | 80 | 1 |
| **Density** | 2.3e-5 | 6.4e-3 | 0.51 |
| **Transitivity** | n/a | 0.28 | 0.74 |
| **Degree assortativity** | n/a | -0.02 | -0.30 |

We first defined a bipartite network (*B*) from the ICPSR Bibliography consisting of publication nodes, dataset nodes, and edges linking publications to the datasets that they cite. Citations are the total number of publications that use data from a study or series. Dataset citations are based on the number of connections shared with publication nodes in the bipartite network (*B*). From network *B*, we projected dataset nodes to create a weighted dataset co-citation network (*S*). Edge weight in *S* indicates the total number of times that a pair of datasets have been used together in publications. We removed low frequency data co-citations from our analysis to focus on datasets that were used together across multiple publications; we removed edges from *S* with a weight less than two, meaning that those datasets were only used together once. This reduced edges by 87% (from 24,942 to 3,208) and nodes by 70% (from 3,363 to 998).

Next, we used a similar process to define a field of research network (*F*) (Cunningham et al., 2022). We gathered supplementary publication metadata for a subset of 44,639 publications in the ICPSR Bibliography (45% of the total) that were available in the Dimensions database (Hook et al., 2018). We retrieved field of research (FoR) codes for each publication. FoR codes consist of 22 high level divisions and their sub-groups (e.g., Curriculum and Pedagogy is a subgroup of Education). We linked FoR codes to ICPSR datasets through their corresponding publications in an unweighted bipartite network (*B*'). We then projected the FoR nodes to create a weighted co-citation network (*F*). In *F*, edges are datasets that are co-cited between fields of research. Because each study could be cited by many different combinations of fields of research, we did not group studies by their series, allowing for the observation of different co-citation patterns in the same series of studies. Edge weight indicates the total number of times a pair of datasets have been used together in publications. We simplified *F* by removing low frequency FoR co-citations, which correspond to edges with a weight less than five.

## Community detection

We applied community detection algorithms to each network as summarized in **Table 4**. Community detection identifies nodes that have a high probability of interacting based on the network structure (Fortunato & Hric, 2016). We selected detection approaches based on the desired representation of communities in each type of network (Lancichinetti & Fortunato, 2009; Yang et al., 2017). We allowed communities to overlap in the dataset co-citation network because we wanted to identify datasets with multiple roles. However, we did not allow overlap in the field of research network because we wanted to find communities defined by members with the strongest ties.

| **Table 4.** Summary of community detection approaches | | | | |
|---|---|---|---|---|
| ***Network*** | **Definition** | **Community detection method** | **Community definition** | **Communities detected** |
| ***S*** | Datasets (studies or series) | k-clique (k=3) | Datasets used in the same paper | 41 |
| ***F*** | Fields of Research (FoR) in papers | Louvain | Fields of research that use the same study-level data | 4 |

We applied a k-clique percolation method to the dataset co-citation network (*S*) using the corresponding implementation from the networkx Python library (Hagberg et al., 2008). A clique is a complete subgraph of a defined size (k), which can be reached from the cliques of the same community through a series of adjacent cliques, meaning that the cliques share k-1 nodes (Palla et al., 2005). Each node may belong to more than one clique, resulting in overlapping communities. We selected a minimum clique size of three and labeled each community with the three most common ICPSR subject terms for all studies in each clique. Subject terms uniformly describe topics covered by the data and are defined by a controlled vocabulary of social science concepts in the ICPSR Subject Thesaurus, which are assigned during data curation.

We then selected an aggregation-based method to represent communities in our field of research network. We applied the Louvain algorithm to the FoR network (*F*) using the corresponding implementation from the Louvain Python library (Hagberg et al., 2008). The algorithm uses modularity to discover communities in large networks by moving nodes locally to create a network aggregation; communities are merged until the resulting modularity of the overall partition can no longer increase (Blondel et al., 2008). This method results in non-overlapping communities that show the most densely-connected fields of research that co-cite ICPSR datasets. The networks (*S*, *F*) were then arranged with a spring layout, which places nodes with high degrees at the center of the graph.

# Results

We used two network measures – centrality and betweenness – to interpret the importance of datasets and fields of research in their respective co-citation networks (Newman, 2003). First, we calculated each node's degree as the number of connections it shares with all other nodes in the network. High degree nodes are prominent in the network because they are highly connected. We also calculated each node's betweenness centrality by measuring all shortest paths passing through a given node. Nodes with high betweenness function as hubs and connect disparate parts of the network.

We also assessed structural features of the network – number of components, assortativity, density, and transitivity – to compare the dataset and field of research co-citation networks (**Table 3)**. The FoR network is connected, meaning that all of its nodes are in the same component, while the other two networks have multiple components or disconnected subgraphs. This suggests that the FoR network is less complex than the dataset co-citation network. Both *S* and *F* exhibit negative degree assortativity, meaning that their nodes are less likely to be connected to nodes in the network with a similar degree value. This pattern is stronger in *F* (-0.30) than in S (-0.02). Finally, networks *B* and *S* have low density (2.3e-5 and 6.4e-3, respectively), while network *F* is far denser (0.51), indicating that *B* and *S* have comparatively fewer edges linking nodes and are not as easily traversed as *F*.

## Dataset co-citations

The dataset co-citation network (*S*) has a periphery of datasets that have been used together only a few times and a denser core of highly connected datasets, which are often used together. **Figure 1** highlights important, central datasets, which are all found in the largest subgraph at the core of the network. We used natural breaks to determine six datasets with high betweenness and degree centrality, which play important roles in the network (**Table 5**).

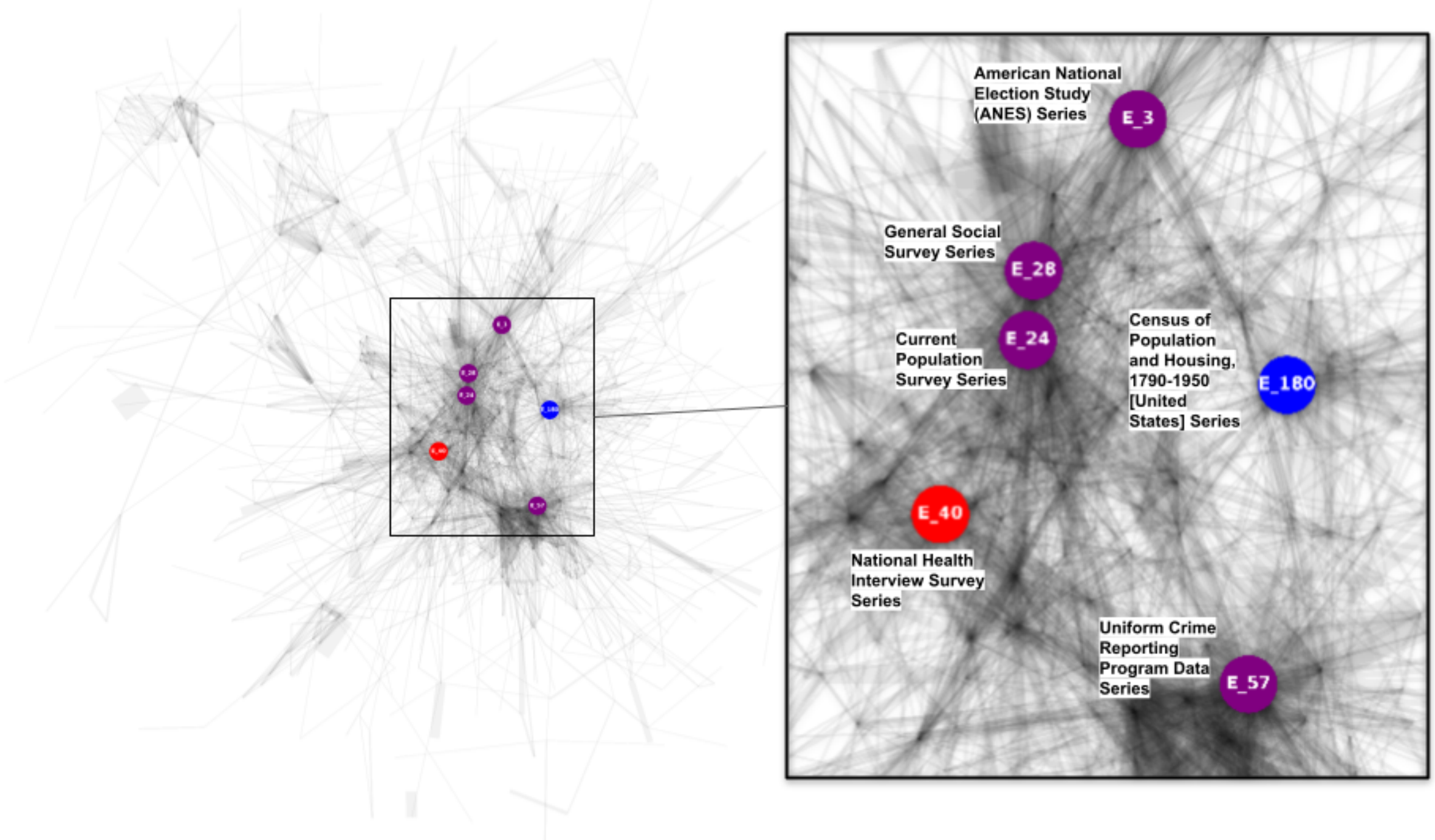

**Figure 1**. Overview of dataset co-citation network featuring datasets functioning as hubs. Inset: High *degree* (red), high *betweenness* (blue), and high *degree* and *betweenness* (purple) nodes.

The important datasets we identified are long-running series made up of multiple studies. Of these, the Uniform Crime Reporting Program Data Series has the highest degree and betweenness. It has been used with 115 other datasets from studies or series across the citation network. The other datasets have strong ties to many other datasets and connect components of the network. Half of these datasets are highly cited with more than 10,000 citations each; the others are less cited, yet play an important role in connecting the network. Finally, the lead investigators for these important datasets include both institutional and non-institutional PIs.

| **Table 5**. Datasets with high betweenness and degree centrality in co-citation network | | | | | |
|---|---|---|---|---|---|
| **Dataset name** | **Investigators** | **Betweenness** | **Degree** | **Studies in series** | **Combined citations** |
| Uniform Crime Reporting Program Data Series | Federal Bureau of Investigation | 0.17 | 115 | 263 | 13,041 |
| General Social Survey Series | National Opinion Research Center; Davis et al. | 0.12 | 113 | 15 | 1,551 |
| American National Election Study (ANES) Series | Miller et al.; National Election Studies | 0.11 | 109 | 92 | 16,771 |
| Current Population Survey Series | U.S. Bureau of the Census | 0.11 | 117 | 296 | 11,012 |
| Census of Population and Housing, 1790-1950 [United States] Series | Haines et al.; U.S. Bureau of the Census | 0.10 | 72 | 2 | 818 |
| National Health Interview Survey Series | National Center for Health Statistics | 0.05 | 80 | 155 | 4,448 |

To find collections of datasets that are often used together in publications, we performed community detection on the dataset co-citation network (**Figure 2**). Not all studies belong to a co-citation community. Only a fraction of datasets in the analysis (N = 632; 63%) belong to cliques of size three or larger; these datasets are often analyzed with at least two additional ICPSR datasets. The datasets that fell out of our analysis were used independently and were not combined with other datasets. We labeled each community with the three most common ICPSR subject terms for all datasets within it. The largest clique has 461 dataset members and is topically broad (e.g., “demographic characteristics, employment, income”) while smaller cliques tend to have narrower focuses (e.g., “terrorism, terrorists, radicalism”).

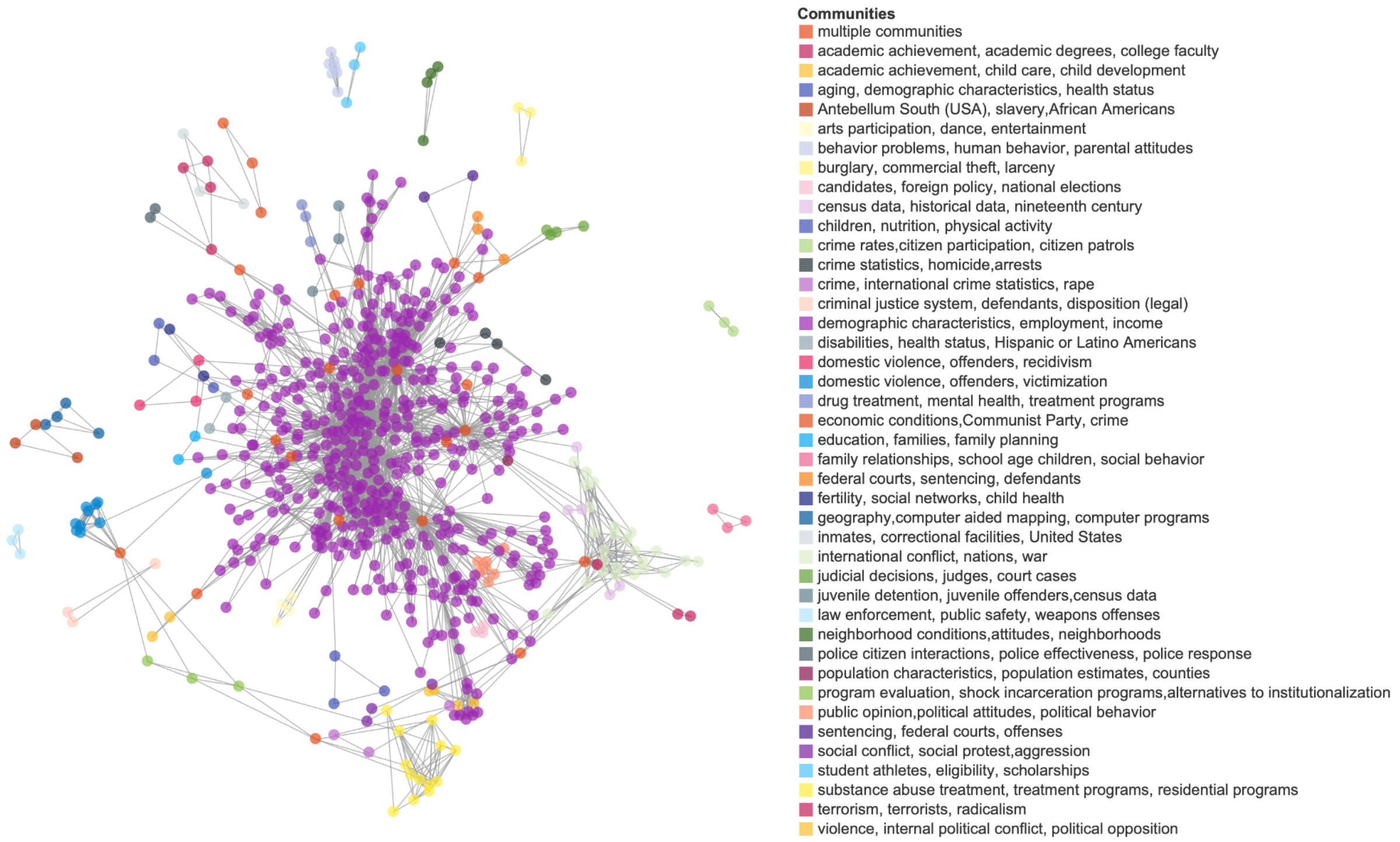

**Figure 2**. Result of community detection (41 communities detected at k=3) with labels generated from the three most frequent subject terms for the datasets in each community. An interactive graph with detailed node information is available in Tableau[7].

We also identified twenty datasets (3% of all nodes in the network) that belong to more than one community, which may facilitate analyses across topics. Of these, we summarized datasets that belong to more than two communities, along with examples of other datasets that they have been co-cited with, and a representative publication that has cited the same data **Table 6**. For example, the Census of Population and Housing, 1790-1950 [United States] Series appears in three different dataset communities. It has been used with other ICPSR datasets to study topics such as industrial development and urbanization in the U.S.; conflict and international trade; and social movements and elections.

**Table 6**. Datasets in more than two communities, their co-cited datasets, and publications

| Dataset | Community label terms | Example of co-cited datasets | Example of citing publication |
|---|---|---|---|
| American National Election Study (ANES) | demographic characteristics, employment, income | National Black Politics Study, [United States], 1993 | Wiegand, A. W. (1999). Differences in public opinion between blacks and whites: A social psychological perspective. University of California, |

[7] https://tinyurl.com/icpsr-datasets

| | | | |
|---|---|---|---|
| Series | | | Santa Cruz. |
| | public opinion, political attitudes, political behavior | Swedish Election Test-Data Series: Swedish Election Study, 1979 | Granberg, D., & Holemberg, S. (1991). Election campaign volatility in Sweden and the United States. Electoral Studies, 10(3), 208-230. |
| | candidates, foreign policy, national elections | American Representation Study, 1958: Candidate and Constituent, Incumbency | Hill, K. Q., & Hurley, P. A. (1979). Mass Participation, Electoral Competitiveness, and Issue-Attitude Agreement Between Congressmen and their Constituents. British Journal of Political Science, 9(4), 507-511. |
| Census of Population and Housing, 1790-1950 [United States] Series | demographic characteristics, employment, income | United States Agriculture Data, 1840 - 2012 | Kitchens, C. T., & Rodgers, L. P. (2020). The Impact of the WWI Agricultural Boom and Bust on Female Opportunity Cost and Fertility (No. w27530). National Bureau of Economic Research. |
| | international conflict, war, nations | Direction of Trade | McKeown, T. J. (1991). A liberal trade order? The long-run pattern of imports to the advanced capitalist states. International Studies Quarterly, 35(2), 151-172. |
| | census data, historical data, nineteenth century | National Samples from the Census of Manufacturing: 1850, 1860, and 1870 | Dobis, E. A. (2016). The Evolution of the American Urban System: History, Hierarchy, and Contagion (Doctoral dissertation, Purdue University). |
| Monitoring of Federal Criminal Sentences Series | demographic characteristics, employment, income | Federal Justice Statistics Program Data Series | Bureau of Justice Statistics (2021). Tribal Crime Data Collection Activities. Technical Report. NCJ 301061, Washington, DC: Bureau of Justice Statistics. |
| | federal courts, sentencing, defendants; | Court Workforce Racial Diversity and Racial Justice in Criminal | Ward, G., Farrell, A., & Rousseau, D. (2009). Does racial balance in workforce representation yield |
| | | Case Outcomes in the United States, 2000-2005 | equal justice? Race relations of sentencing in federal court organizations. Law & Society Review, 43(4), 757-806. |
| | sentencing, federal courts, offenses | Impact of Sentencing Guidelines on the Use of Incarceration in Federal Criminal Courts in the United States, 1984-1990 | Tonry, M. (1991). Mandatory Minimum Penalties and the US Sentencing Commission's Mandatory Guidelines. Federal Sentencing Reporter, 4(3), 129-133. |

## Fields of research

To find fields of research (FoR) that often use the same datasets, we performed community detection on the FoR co-citation network (*F*). Nodes in *F* are color-coded by their parent-level divisions and labeled by their child-level code. We detected four large communities, which are summarized in **Figure 3(a)**. The primary fields of research in each community are: Human Society, Philosophy and Education (Community 0); Economics, Commerce and Management (Community 1); Engineering, Earth and Environment, Information and Computer Science (Community 2); and Medical and Health, Biology (Community 3).

Fields in the center of *F* have more co-citations, meaning that they are highly connected to other fields. The central red frame in **Figure 3(a)** shows the major domains of research that cite ICPSR datasets: Human Society, Philosophy and Education (Community 0). These central domains are consistent with the idea that most items in the ICPSR Bibliography are social science publications. Indeed, social science (e.g., Study of Human Society) and methodological research fields (e.g., Statistics) are found in the core of the network while humanities and other fields (e.g., Creative Writing, Performing Arts) exist mostly on the periphery.

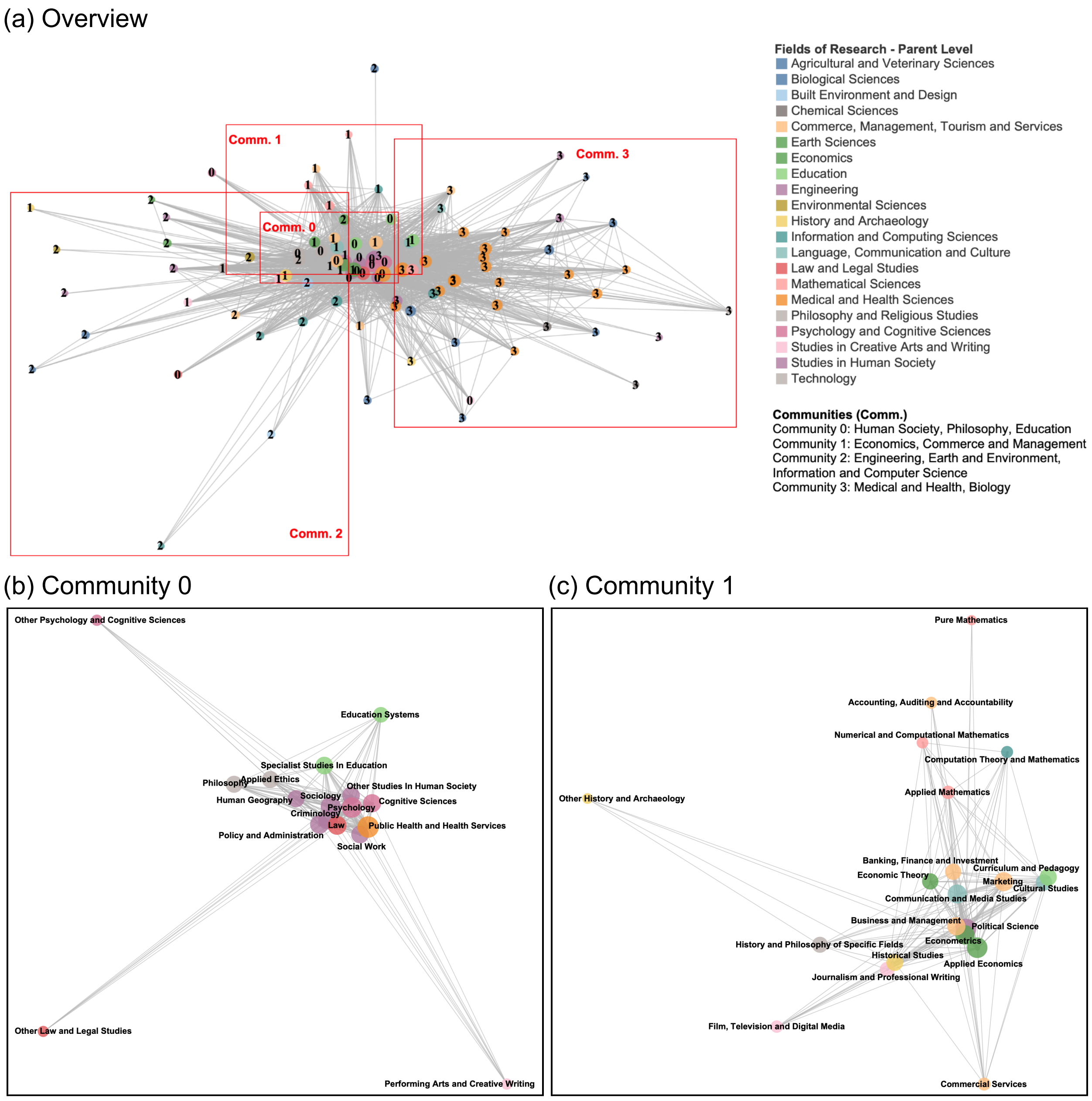
(a) Overview
Comm. 0
Comm. 1
Comm. 2
Comm. 3
Fields of Research - Parent Level
Agricultural and Veterinary Sciences
Biological Sciences
Built Environment and Design
Chemical Sciences
Commerce, Management, Tourism and Services
Earth Sciences
Economics
Education
Engineering
Environmental Sciences
History and Archaeology
Information and Computing Sciences
Language, Communication and Culture
Law and Legal Studies
Mathematical Sciences
Medical and Health Sciences
Philosophy and Religious Studies
Psychology and Cognitive Sciences
Studies in Creative Arts and Writing
Studies in Human Society
Technology
Communities (Comm.)
Community 0: Human Society, Philosophy, Education
Community 1: Economics, Commerce and Management
Community 2: Engineering, Earth and Environment, Information and Computer Science
Community 3: Medical and Health, Biology
(b) Community 0
Other Psychology and Cognitive Sciences
Education Systems
Specialist Studies In Education
Philosophy
Applied Ethics
Other Studies In Human Society
Human Geography
Sociology
Cognitive Sciences
Psychology
Criminology
Law
Public Health and Health Services
Policy and Administration
Social Work
Other Law and Legal Studies
Performing Arts and Creative Writing
(c) Community 1
Pure Mathematics
Accounting, Auditing and Accountability
Numerical and Computational Mathematics
Computation Theory and Mathematics
Other History and Archaeology
Applied Mathematics
Banking, Finance and Investment
Curriculum and Pedagogy
Economic Theory
Marketing
Cultural Studies
Communication and Media Studies
Business and Management
Political Science
History and Philosophy of Specific Fields
Econometrics
Historical Studies
Applied Economics
Journalism and Professional Writing
Film, Television and Digital Media
Commercial Services

(d) Community 2 (e) Community 3

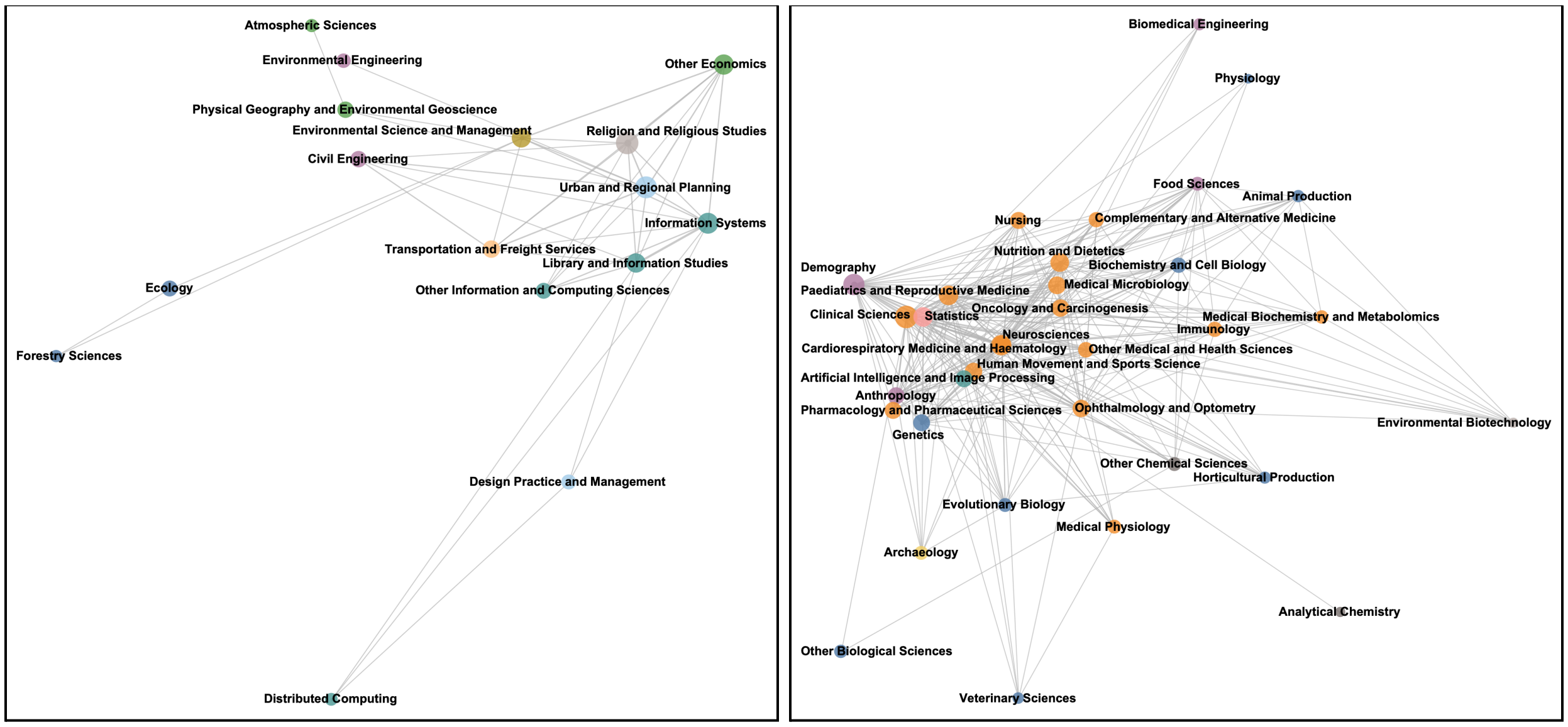

**Figure 3.** Results of community detection in the field of research network (*F*, with nodes connected by edges of size >= 5). The interactive graph with detailed node and edge information in size and Study numbers is available in Tableau[8].

**Figure 3(b)-(e)** shows the composition of each of the four communities in greater detail. We found that the communities tend to divide along disciplinary lines. For example, members within each community are similar, in that they tend to share the same parent-level field of research. For example, “Human Geography” and “Sociology” share the same parent-level field of Human Society and are grouped into the same community (Community 0).

To examine the extent to which similar fields of research use the same datasets, we calculated citation statistics based on network *F*. We consider fields “similar” if they belong to the same parent-level field (e.g., “Civil Engineering” and “Environmental Engineering” are both classified under Engineering) or the same community. We found that similar fields of research co-cite a limited range of datasets. The distribution of the aggregated numbers of datasets for co-citation frequency by parent-level fields of research roughly follows a Poisson distribution with $\lambda = 1$, indicating that as the number of parent-level fields citing the dataset increases, the number of co-citations decreases (**Figure 4(a)**). More than half (2,943 of 5,712) of the datasets in *F* are co-cited by only one community, further suggesting that dataset use tends not to cross community boundaries (**Figure 4(b)**).

[8] https://tinyurl.com/icpsr-fields-of-research

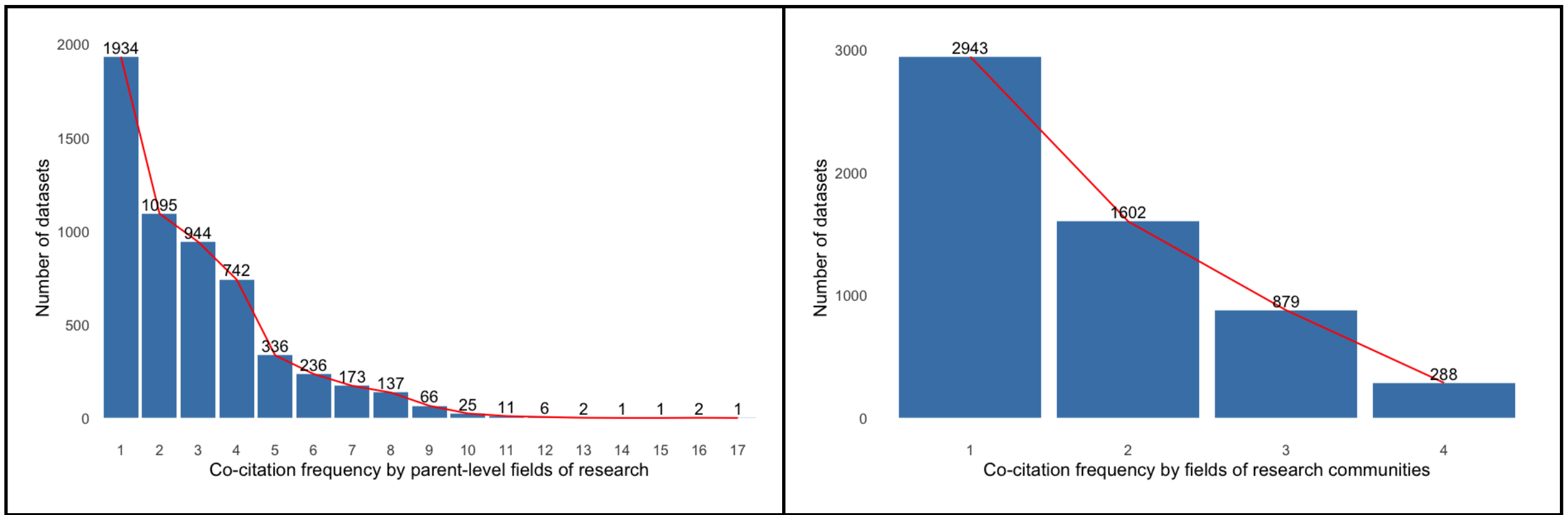

(a) Cross parent-level fields of research

(b) Cross fields of research communities

**Figure 4.** Datasets cited by parent-level fields of research. The *y* axis indicates how many datasets were cited by the number of parent-level fields on the *x* axis. Most datasets are cited by a single parent-level field of research.

We also observed core and periphery structures in the FoR network shown in **Figure 3(a)**. **Table 7(a)** shows examples of fields of research located at the core of each community subgraph. They include a wide range of subfields such as Psychology, Statistics, and Library and Information Studies, which often advance methodological practices and make data-related contributions. These nodes are highly connected to other fields of research and have a much higher degree centrality compared to the average degree of nodes in *F*, which is 9.

Fields of research in the periphery of each community subgraph (**Table 7(b)**) reveal hidden connections among disciplines through the datasets that they co-cite. For example, Archaeology was co-cited by 10 fields – while some of the co-citations are from social science disciplines like Anthropology and Demography, many others are related to biological and physical sciences, including Clinical Sciences, Neurosciences, and Nutrition and Dietetics, which are found in Community 3.

**Table 7**. Examples of non-social science fields of research with core and periphery structures

(a) Fields in the core of each community subgraph

| **Community membership** | **Field of research** | **Number of connected fields of research – degree centrality of nodes** |
|---|---|---|
| 0 | Psychology | 1,833 |
| 0 | Cognitive Sciences | 543 |
| 0 | Law | 1,308 |
| 1 | Applied Mathematics | 26 |
| 2 | Library and Information Studies | 66 |
| 2 | Information Systems | 138 |
| 3 | Statistics | 363 |
| 3 | Artificial Intelligence and Image Processing | 110 |

| (b) Fields in the periphery of each community subgraph | | |
|---|---|---|
| **Community membership** | **Field of research** | **Example of frequently co-cited dataset and corresponding fields of research** |
| 0 | Performing Arts and Creative Writing | “National Crime Victimization Survey: School Crime Supplement, 2011”, co-cited by fields including Policy and Administration, Criminology, Sociology, Specialist Studies in Education, Psychology, Public Health and Health Services, Cognitive Sciences |
| 1 | Curriculum and Pedagogy | “Midlife in the United States (MIDUS 2), 2004-2006”, co-cited by fields including Applied Mathematics, Banking, Finance and Investment, Economic Theory, Communication and Media Studies, Business and Management, Political Science, Econometrics, Applied Economics, Commercial Services |
| 2 | Transportation and Freight Services | “American Time Use Survey (ATUS): Arts Activities, [United States], 2003-2018”, co-cited by fields including Environmental Science and Management, Other Economics, Religion and Religious Studies, Urban and Regional Planning, Information Systems, Library and Information Studies |
| 3 | Archaeology | “National Health and Nutrition Examination Survey III, 1988-1994”, co-cited by fields including Anthropology, Demography, Clinical Sciences, Statistics, Artificial Intelligence and Image Processing, Human Movements and Sports Science, Neurosciences, Nutrition and Dietetics, Other Medical and Health Sciences, Biochemistry and Cell Biology |

# Discussion

In this article we have applied metaphors from the built environment to interpret the hidden research communities that we detected, and labeled the structures *subdivisions* and *crossroads*. These metaphors remind us that these communities of data use have emerged through patterns of interaction in the research landscape and can be reshaped through intentional design. We refer to datasets in research *subdivisions* if they are inward-facing, exclusive, and not well-connected to other datasets or fields. Conversely, we refer to datasets that are often traversed by communities and fields as *crossroads*. We find: 1) research datasets at *crossroads* in the network function as boundary objects by facilitating interdisciplinary research; 2) research datasets in *subdivisions* function as disciplinary resources; and 3) many non-social science fields engage with social science data.

## Subdivisions: disciplinary research community resources

We refer to datasets that serve a single disciplinary community as *subdivisions* because they are inward-facing, exclusive, and are not well-connected. The largest data communities we detected focus on international conflict, substance abuse, victimization, and public opinion polls. Despite the topical breadth of the dataset network (*S*), it partitioned into coherent cliques with a structure better described as a patchwork of subdivisions than a melting pot. By comparison, the FoR network (*F*) had high density and high transitivity, suggesting that its nodes tended to be clustered together. Given its cohesive structure, we partitioned *F* into a small number of meaningful communities.

To understand the communities that function as subdivisions, we drew from a combination of metrics computed for each network, which are summarized in **Table 3**. Overall, the dataset co-citation network (*S*) isn't well-connected. It has low density, low transitivity, is non-assortative based on degree, and contains many components. By comparison, the field of research network (*F*) has a negative degree assortativity, meaning that high-degree fields of research nodes tend to attach to low-degree nodes. The network is not fractured compared with the dataset co-citation network (*S*) and has only one component.

In the dataset network (*S*) shown in **Figure 2**, we found instances of isolated cliques with datasets that were exclusively used together. For example, we detected a clique of three datasets described by the terms "Antebellum South (USA), slavery, slave labor". These datasets ("Southern Farms Study, 1860"; "Mortality in the South, 1850"; and "New Orleans Slave Sale Sample, 1804-1862") have different investigators and were produced for different purposes, yet have been used together numerous times in academic articles. These three studies function like a collection even though ICPSR did not designate them as one (i.e., by naming them a series). In general, the analytic utility of datasets in subdivisions is limited to specific areas of research. The notion of "thematic research collection" – a set of materials on a related theme (Fenlon, 2017; Palmer, 2004) – may be useful for data archives to adopt; finding groups of data used together is one way to identify candidate collections.

We also found examples of cliques that shared topics, yet were disconnected from each other (e.g., "domestic violence, offenders, recidivism" and "domestic violence, offenders, victimization"). While these datasets may be topically similar, researchers have not yet used these data together. Cliques may be exclusive or disjointed for discovery reasons (i.e., researchers outside of the user group are not aware of this data) or their data may be discoverable but unsuitable (e.g., due to variables, geography, or other properties). For example, one community with data about “drug treatment” is composed of studies funded by the United States Department of Health and Human Services, while a separate community of “substance abuse” datasets is funded by the United States Department of Justice. These distinct communities may have stances toward a research topic that are not interoperable and may even conflict.

In the field of research network (*F*) in **Figure 3**, we observed a subdividing tendency and an in-group co-citation pattern for similar fields of research. These patterns of connection suggest that each field of research cites a limited range of ICPSR datasets and supports the idea that ICPSR data use divides along disciplinary boundaries (e.g., social science disciplines like economics and education tend to cite the same datasets, but this is less common across non-social science fields, such as engineering or nursing). Datasets in *subdivisions* have high analytic potential for narrow communities of research; surfacing them and increasing their visibility may also help unlock hidden potential for new uses beyond those narrow communities.

## Crossroads: engagement across research communities

Datasets that facilitate interdisciplinary research are *crossroads* because they are often traversed in connecting communities. For instance, ICPSR is well known for large series datasets (e.g., American National Election Study [ANES]), which attract data users to the archive. We found several of these series in the largest clique (see Table 5), which overlaps with the largest subgraph of the network. These series are well known and have high engagement across multiple research communities. In particular, the ANES Series and the Uniform Crime Reporting Program Data Series are institutionally-funded, highly-cited, and connect a network of researchers who use them.

Prior work found a correlation between datasets with at least one institutional PI and higher data reuse (Hemphill et al., 2022). When we examine data reuse based on citations rather than downloads, however, the relationship between datasets with institutional PIs and reuse is less clear. Some institutional datasets already link multiple data sources into a single dataset and are useful on their own; they may not need to be combined with other datasets to be analytically powerful. Among the crossroads datasets we found, the Census of Population and Housing dataset is unique because the individual investigator who constructed the dataset combined multiple years and data sources into a single dataset, which has been broadly useful across many applications.

In addition to the three connective datasets described in **Table 6**, we found 17 additional datasets that function as crossroads between research communities. Many of these datasets were often used with less cited datasets, explaining the negative associativity observed in network *F*. For example, the less cited “Vietnam Longitudinal Survey, 1995-19987”, is used with the highly cited “India Human Development Survey (IHDS) Series” and “Chitwan Valley [Nepal] Family Study Series” to study education, families, and family planning. Researchers who seek data from a well-known study may traverse the citation network to find complementary datasets from lesser-known studies. While a single data series like the IHDS might meet only some users' needs given its limited geographic coverage, the datasets linked through its connections offer opportunities for comparative analysis.

In the field of research network (*F*), we found two dominant patterns of co-citation, summarized in **Table 7**. Fields in the core of the network are highly connected and operate at an interdisciplinary crossroads; they tend to use more datasets in common with other fields. These fields, like Statistics and Applied Mathematics, are not in the social sciences. Rather, the datasets that they use function as crossroads, activating sites for research convergence. In Community 3 (**Figure 3(e)**) for example, Statistics co-cites many of the same datasets as Biology, Neuroscience, and Medical Sciences. Statistical methods are often applied in data analysis and can advance the development of methodologies in these areas. Fields on the periphery of the network also seem to indicate new forms of engagement with social science data. For example, the field of Transportation and Freight Services uses data from "American Time Use Survey (ATUS): Arts Activities, [United States], 2003-2018," along with Environmental Science and Management, Economics, Religion and Religious Studies, Urban and Regional Planning, Information Systems, Library and Information Studies. Connections between fields on the periphery of each community subgraph appear to maintain weak ties among fields of research (Granovetter, 1973).

## The role of research data in scientific communities

These two structures suggest unique roles for data in scientific communities. Datasets in *subdivisions* and *crossroads* are two types of essential resources supporting social science research; subdivisions may have high disciplinary impact for the specific research domains that use them, while datasets at a crossroads may provide connectivity across domains. For instance, data at *crossroads* enable a kind of "arm's length" cooperative work where the work is loosely coupled, but depends on a "shared information space" that includes common data (Bannon & Schmidt, 1989, p. 361).

While most ICPSR data is used by many disciplines within the bounds of social science, data reuse outside of the social sciences tends to engage with data in two main ways. First, fields such as statistics and artificial intelligence are central in the field of research co-citation network; these fields may reuse social science data to develop new research and analytic methods. Second, fields such as performing arts and creative writing are peripheral in the network; while they tend to reuse ICPSR's data less overall, they may provide novel inroads for "awakening" cross-disciplinary data reuse in new application areas (Hu & Rousseau, 2019).

Identifying hidden communities and their structures within the data citation graph helps us understand how data promotes knowledge production (Buneman et al., 2021; Lowenberg et al., 2019). It's likely that datasets occupying these different structures offer different types of "analytical potential." Palmer, et al. (2011) describe "analytic potential" as "possible analytic contributions for the range of possible user communities" (p. 4), and our method exposes those possible communities and their structure. Research communities are beginning to recognize the importance of contributing to data resources, and the citation graph enables us to assign credit for different kinds of contributions (Alter & Gonzalez, 2018; Cousijn et al., 2019). Naming these different structures provides an accessible, extensible language for discussing the functions of data and assigning credit for their creation. Creating and sharing data that is used widely within one's discipline ought to afford researchers credit among their peers, sometimes for facilitating disciplinary depth – *subdivisions* – and at other times for creating multidisciplinary resources – *crossroads*.

## Limitations and outlook

We relied on the Dimensions database's existing classification scheme for fields of research. This was a pragmatic choice given that codes were assigned at the level of publications rather than journals. However, the granularity of fields of research may be too coarse for interpreting finer, disciplinary patterns of data use within domain archives. Adopting other domain analysis approaches could enhance our understanding of scientific knowledge production (Hjørland & Albrechtsen, 1995). In addition, we could compare the reuse of curated social science data ICPSR to self-archived data (e.g., from the Dutch National Centre of Expertise and Repository for Research Data: DANS).

We were able to identify datasets that served different purposes within scientific communities, but our data does not allow us to comment on how credit for creating different types of data resources ought to be awarded to data creators and providers. Future research should examine the relationship between data creation, reputation, and careers to understand how to recognize data creators' contributions. Because of the different roles they play in connecting and supporting scholarly communities, data creators who produce *subdivisions* or *crossroads* likely deserve different types of credit for their contributions. For instance, creating a dataset that operates as a subdivision should afford data creators substantial credit within their discipline, while creating crossroads may award creators a broader reputation that is less well-recognized within a single discipline. Data creators' academic careers depend on how they receive credit for their work and could impact the types of data resources they create and share.

Our data is essentially a snapshot in time, and they do not enable us to investigate the processes of community formation. However, we are interested in studying these processes, which would explain how social ties, data curation, or other factors shape data citation networks. For example, temporal citation dynamics provide rich insights into the formation of research communities (Chubin, 1976). Extending the idea of "hibernation" to research datasets that have not yet been "awakened" through reuse (Hu & Rousseau, 2019) and detecting bursts of citations following long periods of dormancy would allow us to detect discovery events in the network. Understanding factors associated with novel data reuse would provide evidence to recommend underutilized research data and prioritize funding and credit for specific data curation activities.

# Conclusion

Data citation networks contain hidden information about communities of data users and the roles data play as primary inputs for scientific knowledge production. Through network analysis, we revealed these communities and identified 41 communities of social science datasets, along with four interdisciplinary research communities that use this data. Six important data series connect the co-citation network. Datasets that are used together exclusively form research *subdivisions*, which are valuable data collections for particular disciplines. Other datasets or fields that connect research communities are *crossroads* and have high topical or analytical versatility. Research datasets that are produced for different purposes, such as long-running series data and single-purpose study data, are often used together. Similar fields of research also tend to use the same combinations of data. In conclusion, these findings contribute new ways of seeing scientific communities and make the impacts of research data reuse visible.

# Acknowledgements

Many thanks to Elizabeth Moss and the ICPSR Bibliography staff (Homeyra Banaeefar, Sarah Burchart, and Eszter Palvolgyi-Polyak), David Bleckley, Amy Pienta, and Dharma Akmon of the MICA team at the Inter-university Consortium for Political and Social Research (ICPSR) for their support of this research. We're also grateful to Sagar Kumar and Andrew Schrock for providing feedback on our earlier drafts.

# Author contributions

Sara Lafia: Conceptualization, Methodology, Analysis and interpretation of data, Writing – original draft, Visualization. Lizhou Fan: Methodology, Analysis and interpretation of data, Writing – original draft, Visualization. Andrea Thomer: Conceptualization, Writing – original draft, Supervision, Funding Acquisition. Libby Hemphill: Conceptualization, Methodology, Writing – original draft, Supervision, Funding Acquisition.

# Competing interests

The authors have no competing interests.

# Funding information

This material is based upon work supported by the National Science Foundation under grant 1930645.

# Data availability

Publicly available citation data (derived from the ICPSR Bibliography in February 2022) and code for this article's analysis are available in a Github repository (https://github.com/ICPSR/data-communities). Access to licensed metadata from Dimensions was granted to subscription-only data sources under a license agreement with Digital Science (https://app.dimensions.ai) through the University of Michigan.